\documentclass[prd,preprint,
superscriptaddress,%showpacs,
nofootinbib,tightenlines,
]{revtex4}
\usepackage{epsfig}
\usepackage{slashed}
\usepackage{amsmath,amssymb,bbm,slashed}

\newcommand{\calO}{{\cal O}}

\newcommand{\gA}{\texttt{g}_A}

\newcommand{\logM}{\ln\frac{M}{\mu}}
\newcommand{\logMsq}{\ln^2\frac{M}{\mu}}

\newcommand{\ben}{\begin{displaymath}}
\newcommand{\een}{\end{displaymath}}
\newcommand{\be}{\begin{equation}}
\newcommand{\ee}{\end{equation}}
\newcommand{\bea}{\begin{eqnarray}}
\newcommand{\eea}{\end{eqnarray}}
\begin{document}
\title{Baryon chiral perturbation theory\footnote{Invited talk given at the Third International
Conference on Hadron Physics, TROIA'11, 22 - 25 August 2011, Canakkale, Turkey}}
    \author{S.~Scherer}
    \affiliation{Institut f\"ur Kernphysik, Johannes
    Gutenberg-Universit\"at, D-55099 Mainz, Germany}
\date{December 23, 2011}
\begin{abstract}
   We provide a short introduction to the
one-nucleon sector of chiral perturbation theory and address
the issue of power counting and renormalization.
   We discuss the infrared regularization and the
extended on-mass-shell scheme.
   Both allow for the inclusion of
further degrees of freedom beyond pions and nucleons and the
application to higher-loop calculations.
   As applications we consider the chiral expansion of the nucleon mass to
order ${\cal O}(q^6)$ and the inclusion of vector and axial-vector mesons in the calculation
of nucleon form factors.
   Finally, we address the complex-mass scheme for describing
unstable particles in effective field theory.
\end{abstract}

\maketitle

\section{Introduction}
   Effective field theory (EFT) is a powerful tool in the description of the
strong interactions.
   Generally speaking, an EFT is a low-energy approximation to some underlying,
more fundamental theory.
   The EFT is expressed in terms of a suitable set of effective degrees of freedom,
dominating the phenomena in the low-energy region (see Table \ref{1:table_QCD_EFT}).
   In the context of the strong interactions, the underlying theory is
quantum chromodynamics
(QCD)---a gauge theory with color SU(3) as the gauge group.
   The fundamental degrees of freedom of QCD, quarks and gluons,
carry non-zero color charge.
   Under normal conditions they do not appear as free particles.
   One assumes that any asymptotically observed hadron must be in a
color-singlet state, i.e., a physically observable state is invariant under
SU(3) color transformations.
   The strong increase of the running coupling for large distances possibly
provides a mechanism for the color confinement.
   For the low-energy properties of the strong interactions
another phenomenon is of vital importance.
   The masses of the up and down quarks and, to a lesser extent, also of the
strange quark are sufficiently small that the dynamics of QCD in the chiral
limit, i.e., for massless quarks, is believed to resemble that of the ``real'' world.
   Although a rigorous mathematical proof is not yet available, there are good reasons
to assume that a dynamical spontaneous symmetry breaking (SSB) emerges from
the chiral limit, i.e., the ground state of QCD has a lower symmetry than
the QCD Lagrangian.
   For example, the comparatively small masses of the pseudoscalar octet, the absence of a parity doubling
in the low-energy spectrum of hadrons, and a non-vanishing scalar singlet quark condensate
are indications for SSB in QCD.
   According to the Goldstone theorem, a breakdown of the chiral
$\mbox{SU(3)}_L\times\mbox{SU(3)}_R$ symmetry at the Lagrangian level to
the $\mbox{SU(3)}_V$ symmetry in the ground state implies the existence of
eight massless pseudoscalar Goldstone bosons.
   The finite masses of the pseudoscalar octet of the real world are attributed to
the explicit chiral symmetry breaking by the quark masses in the QCD Lagrangian.
   Both the vanishing of the Goldstone boson masses in the chiral limit and
the vanishing interactions in the zero-energy limit provide a natural
starting point for a derivative and quark-mass expansion.
   The corresponding EFT is called (mesonic) chiral perturbation
theory (ChPT) \cite{Weinberg:1978kz,Gasser:1983yg,Gasser:1984gg}, with the Goldstone bosons as
the relevant effective degrees of freedom
(see, e.g., Refs.~\cite{Scherer:2002tk,Bijnens:2006zp,Bernard:2007zu,Scherer:2009bt,Scherer:2012zzd}
for an introduction and overview).

   Mesonic ChPT may be extended to also include other hadronic degrees of freedom.
   The prerequisite for such an effective field theory program
is (a) a knowledge of the most general effective Lagrangian and (b) an expansion scheme for observables in terms
of a consistent power counting method.
   In the following we will outline some developments of the last decade in devising
renormalization schemes leading to a simple and consistent power counting for the
renormalized diagrams of baryon chiral perturbation theory (BChPT) \cite{Gasser:1987rb}.

\begin{table}[t]
\begin{center}
\renewcommand{\arraystretch}{1.5}
\begin{tabular}{c|c|c}
 &Fundamental theory & Effective field theory\\\hline
 Theoretical framework&QCD&ChPT\\\hline
 Degrees of freedom & Quarks and gluons & Goldstone bosons (+ other hadrons)\\
 \hline
 Parameters & $g_3$ + quark masses & Low-energy coupling constants + quark masses
\end{tabular}
\label{1:table_QCD_EFT} \caption{Comparison of QCD and ChPT.}
\end{center}
\end{table}

\section{Renormalization and Power Counting}
\subsection{Effective Lagrangian and power counting}

  The effective Lagrangian relevant to the one-nucleon sector
consists of the sum of the purely mesonic and $\pi N$ Lagrangians, respectively,
\begin{displaymath}
{\cal L}_{\rm eff}={\cal L}_{\pi}+{\cal L}_{\pi N}={\cal L}_\pi^{(2)}+{\cal L}_\pi^{(4)}+\ldots+{\cal L}_{\pi
N}^{(1)}+{\cal L}_{\pi N}^{(2)}+\ldots,
\end{displaymath}
which are organized in a derivative and quark-mass expansion \cite{Weinberg:1978kz,Gasser:1983yg,Gasser:1987rb}.
   For example, the lowest-order basic Lagrangian ${\cal L}_{\pi N}^{(1)}$,
already expressed in terms of renormalized parameters and fields, is given by
\begin{equation}
\label{LpiN1} {\cal L}_{\pi N}^{(1)}=\bar \Psi \left( i\gamma_\mu
\partial^\mu - m\right) \Psi
-\frac{1}{2}\frac{\texttt{g}_A}{F} \bar \Psi \gamma_\mu \gamma_5 \tau^a \partial^\mu \pi^a \Psi +\ldots,
\end{equation}
where $m$, $\texttt{g}_A$, and $F$ denote the chiral limit of the physical nucleon mass, the axial-vector
coupling constant, and the pion-decay constant, respectively.
   The ellipsis refers to terms containing external fields and
higher powers of the pion fields.
   When studying higher orders in perturbation theory one encounters
ultraviolet divergences.
   As a preliminary step, the loop integrals are regularized,
typically by means of dimensional regularization.
   For example, the simplest dimensionally regularized integral
relevant to ChPT is given by \cite{Scherer:2002tk}
\begin{eqnarray*}
I(M^2,\mu^2,n)&=&\mu^{4-n}\int\frac{\mbox{d}^nk}{(2\pi)^n}\frac{i}{k^2-M^2+i0^+} =\frac{M^2}{16\pi^2}\left(
R+2\ln\frac{M}{\mu}\right)+O(n-4),
\end{eqnarray*}
where $R=\frac{2}{n-4}-[\mbox{ln}(4\pi)+\Gamma'(1)]-1$ approaches infinity as $n\to 4$.
   The 't Hooft parameter $\mu$ is responsible for the fact that the integral has
the same dimension for arbitrary $n$.
   In the process of renormalization the
counter terms are adjusted such that they absorb all the ultraviolet divergences occurring in the calculation of
loop diagrams \cite{Collins:xc}.
   This will be possible, because we include in the Lagrangian all
of the infinite number of interactions allowed by symmetries \cite{Weinberg:1995mt}.
   At the end the regularization is removed by taking the limit
$n\to 4$.
   Moreover, when renormalizing, we still have the freedom of choosing a renormalization
prescription.
   In this context the finite pieces of the renormalized couplings  will be adjusted such that
renormalized diagrams satisfy the following power counting:
   a loop integration in $n$ dimensions counts as $q^n$,
pion and fermion propagators count as $q^{-2}$ and $q^{-1}$, respectively, vertices derived from ${\cal
L}_{\pi}^{(2k)}$ and ${\cal L}_{\pi N}^{(k)}$ count as $q^{2k}$ and $q^k$, respectively.
   Here, $q$ collectively stands for a small quantity such as the pion
   mass, small external four-momenta of the pion, and small external
three-momenta of the nucleon.
   The power counting does not uniquely fix the renormalization scheme,
i.e., there are different renormalization schemes leading to the above specified power counting.

\subsection{Example: One-loop contribution to the nucleon mass}

\begin{figure}[t]
\begin{center}
\epsfig{file=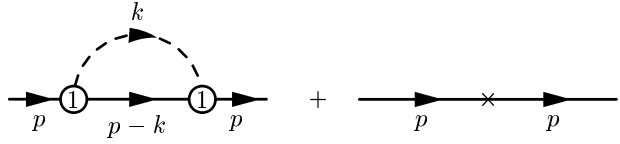,width=0.5\textwidth} \caption{\label{4:2:3:ren_diag} Renormalized one-loop self-energy
diagram. The number 1 in the interaction blobs refers to ${\cal L}_{\pi N}^{(1)}$.
The cross generically denotes counter-term contributions.
}
\end{center}
\end{figure}

   In the mesonic sector, the combination of dimensional regularization and
the modified minimal subtraction scheme $\widetilde{\mbox{MS}}$ leads to a straightforward correspondence between
the chiral and loop expansions \cite{Gasser:1983yg,Gasser:1984gg}.
   By discussing the one-loop contribution of Fig.~\ref{4:2:3:ren_diag}
to the nucleon self energy, we will see that this correspondence, at first sight, seems to be lost in the
baryonic sector.
   According to the power counting specified above, after renormalization,
we would like to have the order $D=n\cdot 1-2\cdot 1-1\cdot 1+1\cdot 2=n-1.$
   An explicit calculation yields \cite{Scherer:2012zzd}
\begin{displaymath}
\Sigma_{\rm loop}= -\frac{3 \texttt{g}_{A}^2}{4 F^2}\left\{
(\slashed{p}+m)I_N+M^2(\slashed{p}+m)I_{N\pi}-\frac{(p^2-m^2)\slashed{p}}{2p^2}[
(p^2-m^2+M^2)I_{N\pi}+I_N-I_\pi]\right\},
\end{displaymath}
where $M^2=2B\hat m$ is the lowest-order expression for the squared pion mass in terms of the low-energy coupling
constant $B$ and the average light-quark mass $\hat m$ \cite{Gasser:1983yg}.
   The relevant loop integrals are defined as
\begin{eqnarray}
\label{Ipi}
I_\pi&=& \mu^{4-n}\int\frac{\mbox{d}^nk}{(2\pi)^n}\frac{i}{k^2-M^2+i0^+},\\
\label{IN}
I_N&=& \mu^{4-n}\int\frac{\mbox{d}^nk}{(2\pi)^n}\frac{i}{k^2-m^2+i0^+},\\
\label{INpi} I_{N\pi}&=&\mu^{4-n}\int\frac{\mbox{d}^nk}{(2\pi)^n}
\frac{i}{[(k-p)^2-m^2+i0^+]}\frac{1}{k^2-M^2+i0^+}.
\end{eqnarray}
   The application of the $\widetilde{\rm MS}$ renormalization scheme of ChPT
\cite{Gasser:1983yg,Gasser:1987rb}---indicated by ``r''---yields
\begin{displaymath}
\Sigma_{\rm loop}^r=-\frac{3 \texttt{g}_{Ar}^2}{4 F^2} \left[M^2(\slashed{p}+m)I_{N\pi}^r+\ldots\right].
\end{displaymath}
   The expansion of $I_{N\pi}^r$ is given by
\begin{displaymath}
\label{4:2:3:INpExp}
    I_{N\pi}^r=\frac{1}{16\pi^2}\left(-1+\frac{\pi M}{m}+\ldots\right),
\end{displaymath}
resulting in $\Sigma_{\rm loop}^r={\cal O}(q^2)$.
   In other words, the $\widetilde{\rm MS}$-renormalized result does not
produce the desired low-energy behavior which, for a long time, was interpreted as the absence of a systematic
power counting in the relativistic formulation of ChPT.

   The expression for the nucleon mass $m_N$ is obtained by solving
the equation
\begin{displaymath}
\label{4:2:3:MassDef} m_N-m-\Sigma(m_N)=0,
\end{displaymath}
from which we obtain for the nucleon mass in the $\widetilde{\rm MS}$ scheme \cite{Gasser:1987rb},
\begin{equation}
\label{4:2:3:MassMStilde}
    m_N=m-4c_{1r}M^2+
    \frac{3\texttt{g}_{Ar}^2M^2}{32\pi^2F^2_r}m
    -\frac{3\texttt{g}_{Ar}^2M^3}{32\pi F^2_r}.
\end{equation}
   At ${\cal O}(q^2)$, Eq.~(\ref{4:2:3:MassMStilde}) contains besides the undesired loop
contribution proportional to $M^2$ the tree-level contribution $-4c_{1r}M^2$ from the next-to-leading-order
Lagrangian ${\cal L}_{\pi N}^{(2)}$.

   The solution to the power-counting problem is the observation
that the term violating the power counting, namely, the third on the right-hand side of
Eq.~(\ref{4:2:3:MassMStilde}), is \emph{analytic} in the quark mass and can thus be absorbed in counter terms.
   In addition to the $\widetilde{\rm MS}$ scheme we have to perform an additional
{\em finite} renormalization.
   For that purpose we rewrite
\begin{equation}
\label{4:2:3:cRenorm}
    c_{1r}=c_1+\delta c_1,\quad \delta c_1 =\frac{3 m {\texttt g}_A^2}{128 \pi^2 F^2}+\ldots
\end{equation}
in Eq.~(\ref{4:2:3:MassMStilde}) which then gives the final result for the nucleon mass at ${\cal O}(q^3)$:
\begin{equation}
\label{4:2:3:MassFinal}
    m_N=m-4c_{1}M^2
    -\frac{3\texttt{g}_{A}^2M^3}{32\pi F^2}.
\end{equation}
   We have thus seen that the validity of a power-counting scheme is intimately
connected with a suitable renormalization condition.
   In the case of the nucleon mass, the $\widetilde{\rm MS}$ scheme alone does not
suffice to bring about a consistent power counting.

\subsection{Infrared regularization and extended on-mass-shell scheme}
   Several methods have been suggested to obtain a consistent power
counting in a manifestly Lorentz-invariant approach.
   We will illustrate the underlying ideas in terms of a typical one-loop integral
in the chiral limit,
\begin{displaymath}
H(p^2,m^2;n)= \int \frac{{\mbox d}^n k}{(2\pi)^n} \frac{i}{[(k-p)^2-m^2+i0^+][k^2+i0^+]},
\end{displaymath}
where $\Delta=(p^2-m^2)/m^2={\cal O}(q)$ is a small quantity.
   Applying the dimensional counting analysis of
Ref.~\cite{Gegelia:1994zz}, the result of the integration is of the form
\begin{displaymath}
H\sim F(n,\Delta)+\Delta^{n-3}G(n,\Delta),
\end{displaymath}
where $F$ and $G$ are hypergeometric functions which are analytic for $|\Delta|<1$ for any $n$.

   In the infrared regularization of Becher and Leutwyler \cite{Becher:1999he}
one makes use of the Feynman parameterization
\begin{displaymath}
\frac{1}{ab}=\int_0^1 \mbox{d}z\frac{1}{[az+b(1-z)]^2}
\end{displaymath}
with $a=(k-p)^2-m^2+i0^+$ and $b=k^2+i0^+$.
   The resulting integral over the Feynman parameter $z$ is then rewritten as
\begin{eqnarray*}
H=\int_0^1 \mbox{d}z \ldots &=& \int_0^\infty \mbox{d}z \ldots
- \int_1^\infty \mbox{d}z \ldots,\\
\end{eqnarray*}
where the first, so-called infrared (singular) integral satisfies the power counting, while the remainder
violates power counting but turns out to be regular and can thus be absorbed in counter terms.

   The central idea of the extended on-mass-shell (EOMS)
scheme \cite{Gegelia:1999gf,Fuchs:2003qc} consists of subtracting those terms which violate the
power counting as $n\to 4$.
   Since the terms violating the power counting are analytic in small
quantities, they can be absorbed by counter-term contributions.
   In the present case, we want the renormalized integral to be of
the order $D=n-1-2=n-3$.
   To that end one first expands the integrand in
small quantities and subtracts those integrated terms whose order is smaller than suggested by the power
counting.
   The corresponding subtraction term reads
\begin{displaymath}
H^{\rm subtr}=\int \frac{\mbox{d}^n k}{(2\pi)^n}\left. \frac{i}{[k^2-2p\cdot k +i0^+][k^2+i0^+]}\right|_{p^2=m^2}
\end{displaymath}
and the renormalized integral is written as $ H^R=H-H^{\rm subtr}={\cal O}(q) $ as $n\to 4$.

\subsection{Remarks}

   Using a suitable renormalization condition, one obtains a consistent power counting in manifestly
Lorentz-invariant baryon ChPT including, e.g., (axial) vector mesons \cite{Fuchs:2003sh} or the $\Delta(1232)$
resonance \cite{Hacker:2005fh} as explicit degrees of freedom.
   The infrared regularization of Becher and
Leutwyler \cite{Becher:1999he} has been reformulated in a form analogous to the EOMS renormalization
\cite{Schindler:2003xv}.
   The application of both infrared and extended on-mass-shell renormalization schemes to
multi-loop diagrams was explicitly demonstrated by means of a two-loop self-energy diagram
\cite{Schindler:2003je}.
   A treatment of unstable particles such as the rho meson or the Roper resonance is possible
in terms of the complex-mass scheme (CMS) \cite{Djukanovic:2009zn,Djukanovic:2009gt}.

\section{Applications}
\label{section_applications}
   In the following we will illustrate a few selected applications
of the manifestly Lorentz-invariant framework to the one-nucleon sector.

\subsection{Nucleon mass to ${\cal O}(q^6)$}
\label{subsec:NucMass}

   The nucleon mass $m_N$ provides a good testing ground for applications of BChPT,
as it does not depend on any momentum transfers and the chiral expansion therefore
corresponds to an expansion in the quark masses.
   For this reason, the calculation of the nucleon mass has been performed in various renormalization
schemes \cite{Gasser:1987rb,Bernard:1992qa,McGovern:1998tm,Becher:1999he,Fuchs:2003qc}.
   With the exception of Ref.~\cite{Gasser:1987rb}, these schemes have in common that they
establish the connection between the chiral and the loop expansions, analogous to the mesonic sector.
   A calculation that only includes one-loop diagrams can be performed up to and including ${\cal O}(q^4)$.
   The general form of the chiral expansion is given by
\begin{equation}
\label{eq:NucMass4} m_N = m + k_1 M^2 + k_2 M^3 + k_3 M^4 \logM +k_4 M^4 + \ldots,
\end{equation}
where $M^2=2B\hat m$ is the lowest-order expression for the squared pion mass, the ellipsis stands for
higher-order terms, and $\mu$ is a renormalization scale. As an example, in the EOMS scheme the expressions for
the $k_i$ are given by \cite{Fuchs:2003qc}
\begin{equation}
\begin{split}
k_1 &= -4c_1,\quad k_2 = -\frac{3\gA^2}{32\pi F^2} , \quad
k_3 = \frac{3}{32\pi^2 F^2} \left( 8c_1-c_2-4c_3-\frac{\gA^2}{m} \right),\\
k_4 &= \frac{3\gA^2}{32\pi^2 F^2 m}(1+4c_1 m) + \frac{3}{128\pi^2 F^2}c_2-2(8e_{38}+e_{115}+e_{116}).
\end{split}
\end{equation}
Here, the $c_i$ and $e_j$ are low-energy constants (LECs) of the second- and fourth-order
baryonic Lagrangians, respectively.
   The expression of Eq.~(\ref{eq:NucMass4}) together with estimates of the various LECs
\cite{Becher:2001hv} was used in Ref.~\cite{Fuchs:2003kq} to determine the nucleon mass in the chiral limit,
\begin{align}
m &= m_N -\Delta m \notag\\
&= (938.3 - 74.8+15.3+4.7+1.6-2.3)\, \text{MeV}\\
&= 882.8\,\text{MeV},\notag
\end{align}
i.e., $\Delta m = 55.5\,\text{MeV}$.
\begin{figure}[t]
\begin{minipage}{0.4\textwidth}
\begin{center}
\includegraphics[width=\textwidth]{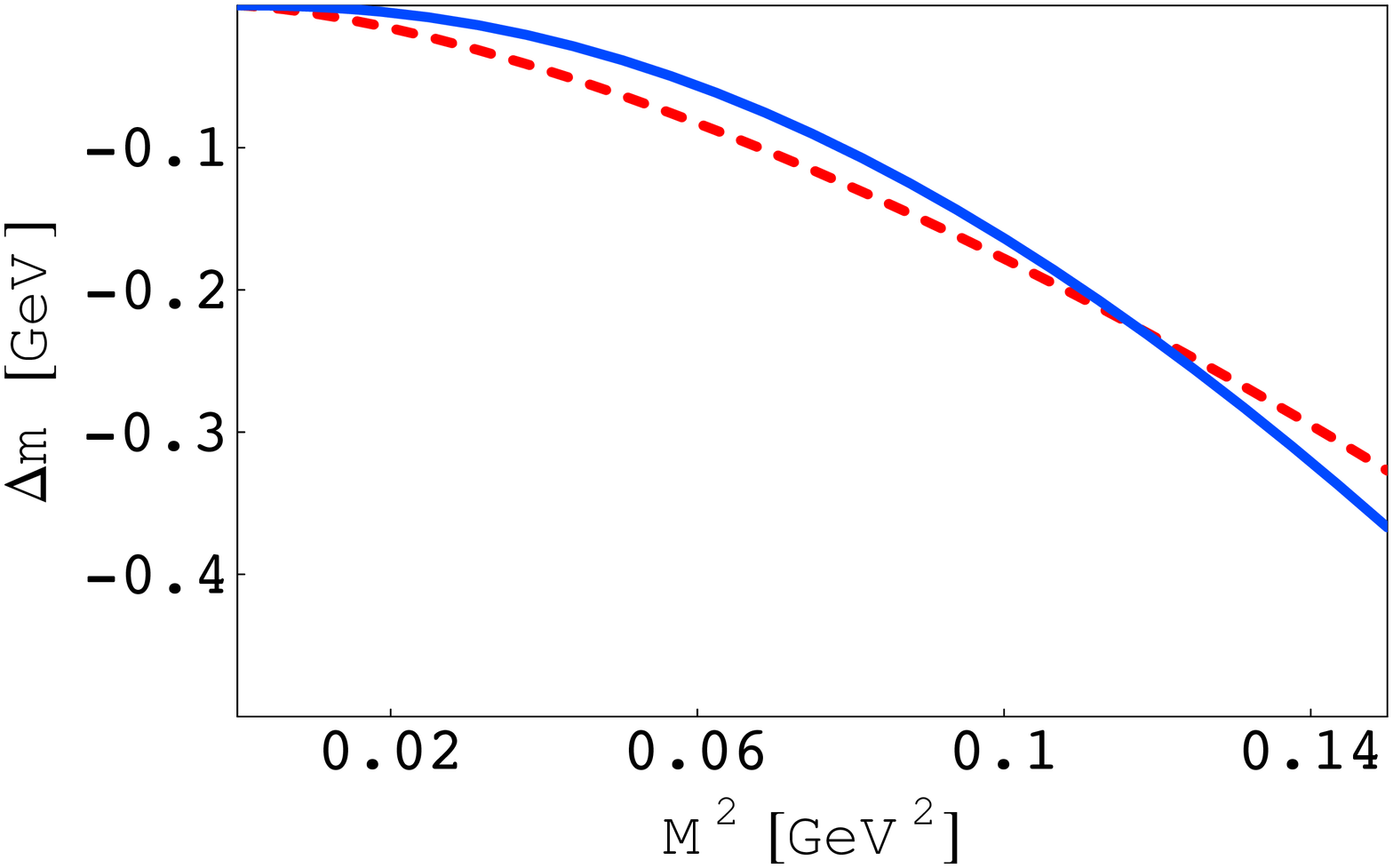}
\end{center}
\end{minipage}
\hspace{0.1\textwidth}
\begin{minipage}{0.4\textwidth}
\begin{center}
\includegraphics[width=\textwidth]{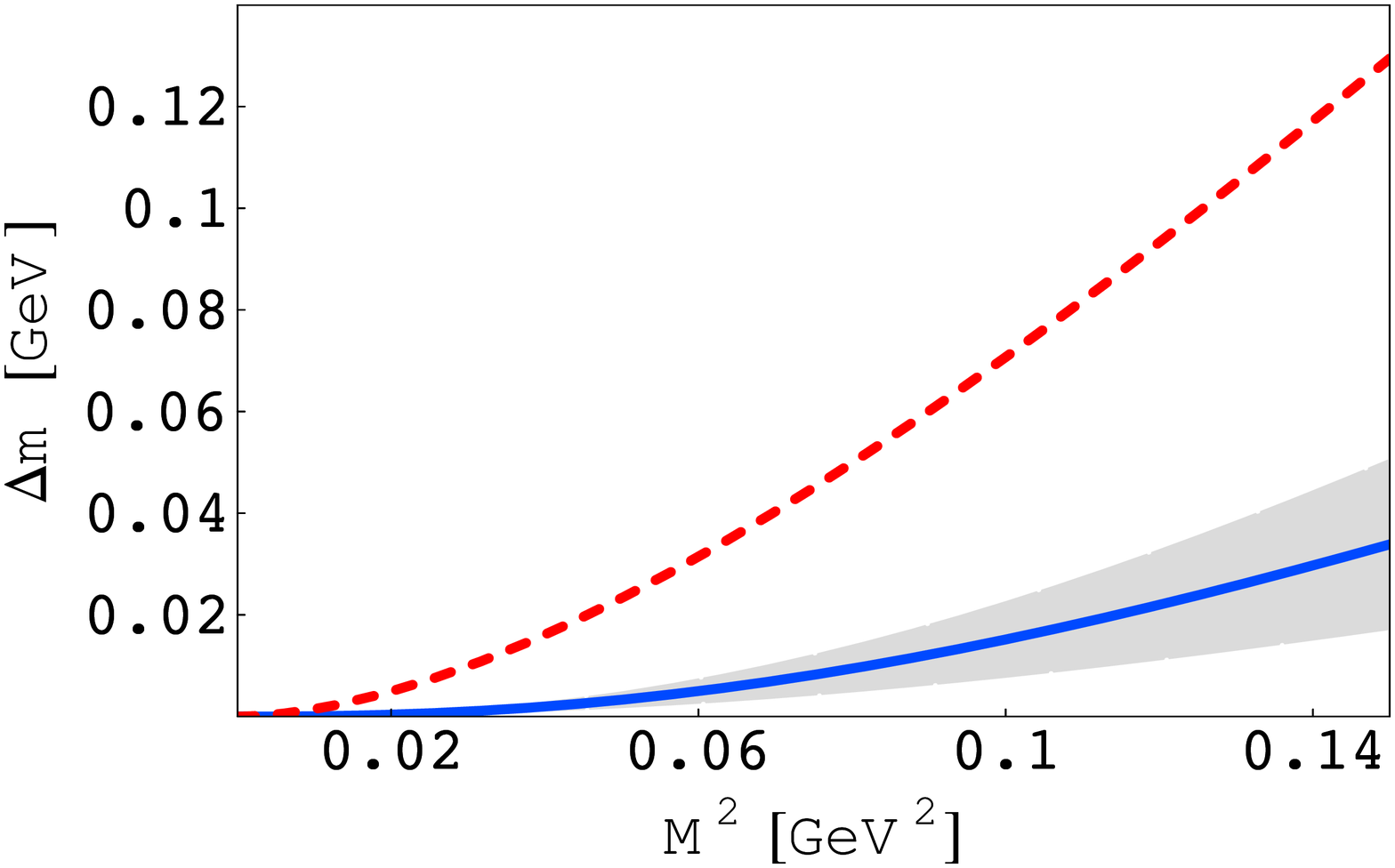}
\end{center}
\end{minipage}
\caption{Pion-mass dependence of terms contributing to the chiral expansion of the nucleon mass. Left panel: The
solid line corresponds to $ k_5 M^5\logM$, the dashed line to $k_2M^3$. Right panel: The solid line corresponds
to $ k_7 M^6\logMsq$, the dashed line to $k_3M^4\logM$. The grey band indicates an  error estimate (see
Ref.~\cite{Schindler:2007dr}).\label{fig:NucMass}}
\end{figure}
Contributions to the nucleon mass at $\calO (q^5)$, i.e., including two-loop diagrams, were first considered in
Ref.~\cite{McGovern:1998tm}, and a complete calculation to  $\calO (q^6)$ was performed in
Refs.~\cite{Schindler:2006ha,Schindler:2007dr}. The higher-order contributions take the form
\begin{equation}
\label{eq:NucMass6}
\begin{split}
m_N &= m + k_1 M^2 + k_2 M^3 + k_3 M^4 \logM +k_4 M^4 \\
&\quad + k_5 M^5\logM + k_6 M^5 + k_7 M^6 \logMsq + k_8 M^6 \logM + k_9 M^6.
\end{split}
\end{equation}
Since various so-far undetermined LECs enter the expressions for some of the higher-order $k_i$ it is not
possible to give an accurate estimate of all terms in Eq.~(\ref{eq:NucMass6}). However, the fifth-order
contribution $ k_5 M^5\logM$ is found to be $ k_5 M^5\ln\frac{M}{m_N} = -4.8\,\text{MeV}$ at the physical pion
mass with $\mu=m_N$, while $ k_6 M^5 = 3.7 \,\text{MeV}$ or  $k_6 M^5 = -7.6\,\text{MeV}$ depending on the choice
of the third-order LEC $d_{16}$ \cite{Schindler:2007dr}. Equation~(\ref{eq:NucMass6}) can also be used to examine
the pion-mass dependence of the nucleon mass, which plays an important role in the extrapolation of lattice QCD
to physical quark masses. Figure~\ref{fig:NucMass} shows the comparison of various terms in
Eq.~(\ref{eq:NucMass6}) as a function of the squared pion mass. While the right panel shows the expected
suppression of the higher-order term over the whole pion-mass range, the left panel indicates that the term $ k_5
M^5\logM$ becomes as large as $k_2M^3$ for a pion mass of roughly $M\sim 360\,\text{MeV}$. While this is not a
reliable prediction of the behavior of higher-order contributions since only the leading nonanalytic parts are
considered, the pion mass range at which the power counting is no longer applicable agrees with the estimates
found in Refs.~\cite{Meissner:2005ba,Djukanovic:2006xc}.

\subsection{Probing the convergence of perturbative series}
   The issue of the convergence of perturbative calculations
is presently of great interest in the context of chiral extrapolations of baryon properties (see, e.g.,
Refs.~\cite{Leinweber:2003dg,Procura:2003ig,Beane:2004ks}).
   A possibility of exploring the convergence of perturbative series
consists of summing up certain sets of an infinite number of diagrams by solving integral equations exactly and
comparing the solutions with the perturbative contributions \cite{Djukanovic:2006xc}.
\begin{figure}[t]
\begin{center}
\epsfig{file=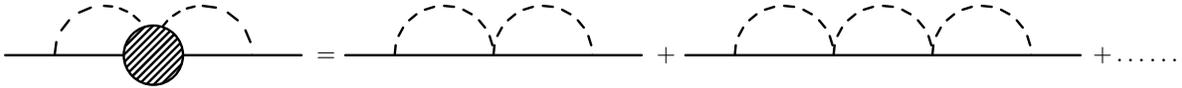, width=\textwidth}
\end{center}
\caption{Iterated contribution to the nucleon self energy.\label{Nse:figdiagrams}}
\end{figure}
   Figure \ref{Nse:figdiagrams} shows a graphical representation
of an iterated contribution to the nucleon self energy originating from the Weinberg-Tomozawa term in the $\pi N$
scattering amplitude.
   The result is of the form \cite{Djukanovic:2006xc}
\begin{equation}
\delta m = -\frac{3\texttt{g}_{A}^2}{4F^2} \frac{N}{D}, \label{contributioninmass}
\end{equation}
where $N$ and $D$ are closed expressions in terms of the loop functions of Eqs.~(\ref{Ipi}) - (\ref{INpi}).
   By expanding Eq.~(\ref{contributioninmass}) in
powers of $1/F^2$ one can identify the contributions of each diagram separately.
   Using the IR renormalization scheme and substituting $m=883$ MeV,
$m_N=938.3$ MeV, $F=92.4$ MeV, $\texttt{g}_{A}=1.267$ and $M=139.6$ MeV one obtains
\begin{equation}
\delta m = -0.00233530\,{\rm MeV}=\left( -0.00230219-0.00003305 - 0.00000007 +\ldots\right)\,{\rm MeV}\,.
\label{dmBL}
\end{equation}
   The first term in the perturbative expansion reproduces the non-perturbative result
well and the higher-order corrections are clearly suppressed.
   Figure \ref{Nseplot:fig} shows $\delta m$ of
Eq.~(\ref{contributioninmass}) together with the leading contribution (first diagram in
Fig.~\ref{Nse:figdiagrams}) and the leading non-analytic correction to the nucleon mass $\delta
m_3=-3g_A^2M^3/(32\pi F^2)$ \cite{Gasser:1987rb} as functions of $M$. As can be seen from this figure, up
to $M\sim 500$ MeV the non-perturbative sum of higher-order corrections is suppressed in comparison with the
$\delta m_3$ term. Also, the leading higher-order contribution reproduces the non-perturbative result quite well.
On the other hand, for $M\gtrsim 600$ MeV the higher-order contributions are no longer suppressed in comparison
with $\delta m_3$.

\begin{figure}
\begin{center}
\epsfig{file=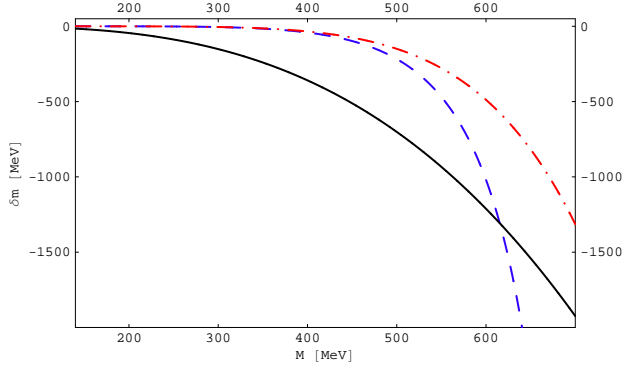, width=0.5\textwidth} \caption[]{\label{Nseplot:fig} Contributions to the nucleon
mass as functions of $M$. Solid line: ${\cal O}(q^3)$ contribution, dashed line: $\delta m$ of
Eq.~(\ref{contributioninmass}); dashed-dotted line: two-loop diagram of Fig.~\ref{Nse:figdiagrams}.}
\end{center}
\end{figure}

\subsection{Electromagnetic form factors}
\label{subsec:FFVM}
   Imposing the relevant symmetries such as translational invariance,
Lorentz covariance, the discrete symmetries, and current conservation, the nucleon matrix element of the
electromagnetic current operator $J^\mu(x)$ can be parameterized in terms of two form factors,
\begin{equation}
\label{H1:emff:empar} \langle N(p')|J^{\mu}(0)|N(p)\rangle= \bar{u}(p')\left[F_1^N(Q^2)\gamma^\mu
+i\frac{\sigma^{\mu\nu}q_\nu}{2m_p}F_2^N(Q^2) \right]u(p),\quad N=p,n,
\end{equation}
   where $q=p'-p$, $Q^2=-q^2$, and $m_p$ is the proton mass.
   At $Q^2=0$, the so-called Dirac and Pauli form factors $F_1$ and
$F_2$ reduce to the charge and anomalous magnetic moment in units of the elementary charge and the nuclear
magneton $e/(2m_p)$, respectively,
\begin{displaymath}
F_1^{p}(0)=1,\quad F_1^{n}(0)=0,\quad F_2^{p}(0)=1.793,\quad F_2^{n}(0)=-1.913.
\end{displaymath}
   The Sachs form factors $G_E$ and $G_M$ are linear combinations of $F_1$ and
$F_2$,
\begin{displaymath}
G_E^N(Q^2)=F_1^N(Q^2)-\frac{Q^2}{4m_p^2}F_2^N(Q^2),\quad G_M^N(Q^2)=F_1^N(Q^2)+F_2^N(Q^2), \quad N=p,n.
\end{displaymath}

\begin{figure}[tb]
\begin{center}
\epsfig{file=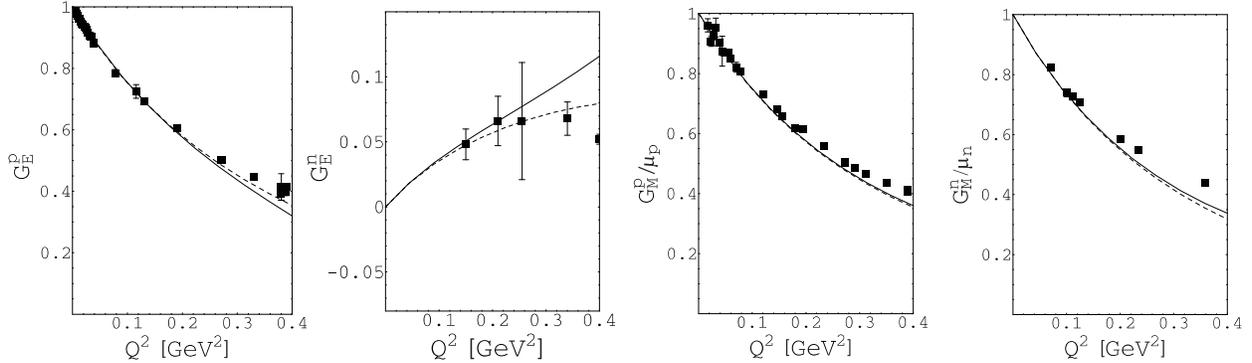,width=\textwidth} \caption{\label{H1:emff:G:neu} The Sachs form factors of the
nucleon in manifestly Lorentz-invariant chiral perturbation theory at ${\cal O}(q^4)$ including vector mesons as
explicit degrees of freedom. Full lines: results in the extended on-mass-shell scheme; dashed lines: results in
infrared regularization. The experimental data are taken from Ref.\ \cite{Friedrich:2003iz}.}
\end{center}
\end{figure}

   Calculations in Lorentz-invariant baryon ChPT up to
fourth order fail to describe the proton and nucleon form factors for momentum transfers beyond $Q^2\sim 0.1\,
\mbox{GeV}^2$ \cite{Kubis:2000zd,Fuchs:2003ir}.
   In Ref.\ \cite{Kubis:2000zd} it was shown that the
inclusion of vector mesons can result in the re-summation of important higher-order contributions.
   In Ref.\ \cite{Schindler:2005ke} the electromagnetic form factors of the
nucleon up to fourth order have been calculated in manifestly Lorentz-invariant ChPT with vector mesons as
explicit degrees of freedom.
   A systematic power counting for the renormalized diagrams has been
implemented using both the extended on-mass-shell renormalization scheme and the reformulated version of infrared
regularization.
    The inclusion of vector mesons results in a
considerably improved description of the form factors (see Fig.~\ref{H1:emff:G:neu}).
    The most dominant contributions come from
tree-level diagrams, while loop corrections with internal vector meson lines are small \cite{Schindler:2005ke}.

\subsection{Axial and induced pseudoscalar form factors}
   Assuming isospin symmetry, the most general
parametrization of the isovector axial-vector current evaluated between one-nucleon states is given by
\begin{equation}\label{H1_axff_FFDef}
\langle N(p')| A^{\mu,a}(0) |N(p) \rangle = \bar{u}(p') \left[\gamma^\mu\gamma_5 G_A(Q^2)
+\frac{q^\mu}{2m_N}\gamma_5 G_P(Q^2) \right] \frac{\tau^a}{2}u(p),
\end{equation}
where $q=p'-p$, $Q^2=-q^2$, and $m_N$ denotes the nucleon mass.
   $G_A(Q^2)$ is called the axial form factor and
$G_P(Q^2)$ is the induced pseudoscalar form factor.
     The value of the axial form factor at zero momentum transfer is defined as
the axial-vector coupling constant, $g_A=G_A(Q^2=0) =1.2694(28)$, and is quite precisely determined from neutron
beta decay.
   The $Q^2$ dependence of the axial form factor can be obtained
either through neutrino scattering or pion electroproduction.
   The second method makes use of the so-called Adler-Gilman relation
\cite{Adler:1966gd} which provides a chiral Ward identity establishing a connection between charged pion
electroproduction at threshold and the isovector axial-vector current evaluated between single-nucleon states
(see, e.g., Ref.\ \cite{Bernard:2001rs,Fuchs:2003vw} for more details).
   The induced pseudoscalar form factor $G_P(Q^2)$ has been investigated in
ordinary and radiative muon capture as well as pion electroproduction (see Ref.\ \cite{Gorringe:2002xx} for a
review).

\begin{figure}[tb]
\begin{minipage}[b]{0.3\textwidth}
\includegraphics[width=\textwidth]{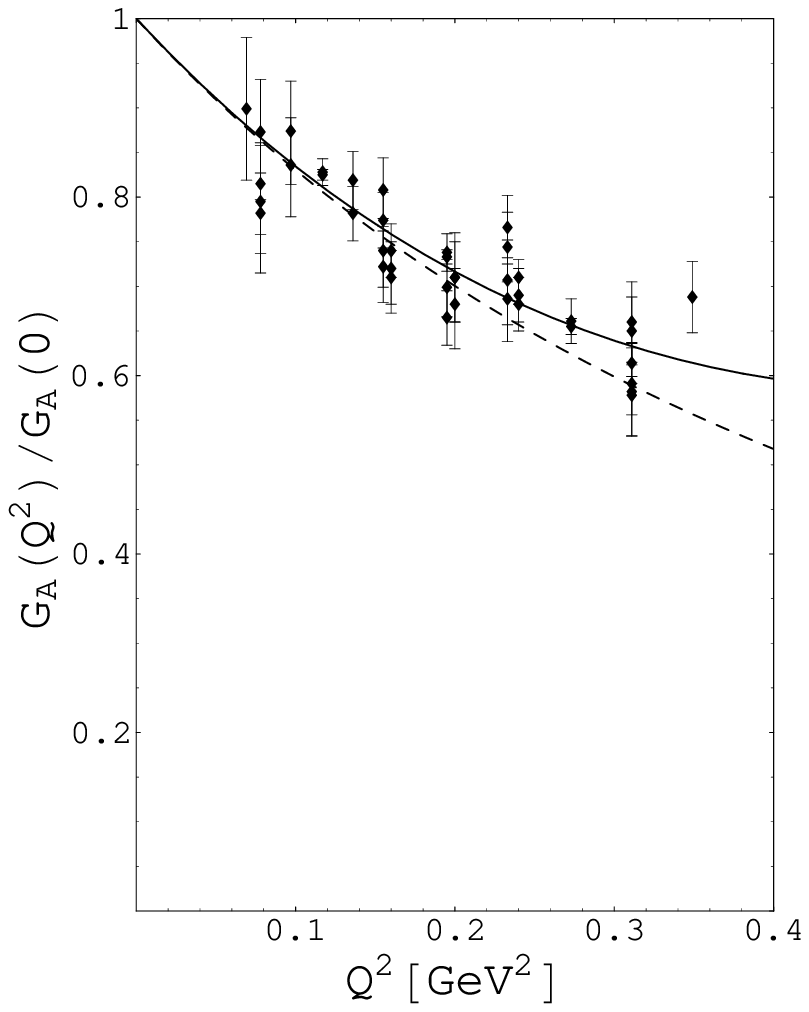}
\end{minipage}
\hspace{2em}
\begin{minipage}[b]{0.5\textwidth}
\includegraphics[width=\textwidth]{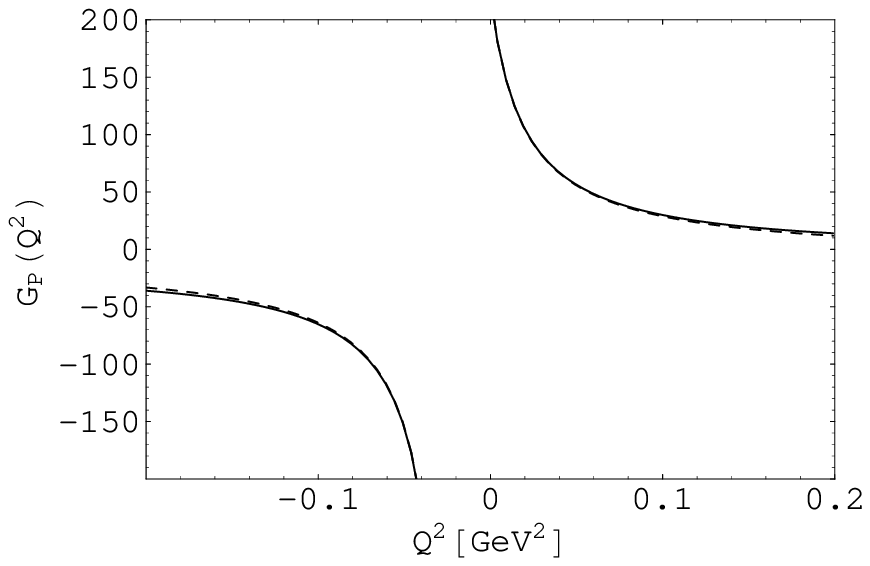}
\end{minipage}
\caption{\label{H1_axff_GAwith}Left panel: Axial form factor $G_A$ in manifestly Lorentz-invariant ChPT at ${\cal
O}(q^4)$ including the axial-vector meson $a_1(1260)$ explicitly. Full line: result in infrared renormalization,
dashed line: dipole parametrization. The experimental data are taken from Ref.\ \cite{Bernard:2001rs}.
 Right
panel:
   The induced pseudoscalar form factor $G_P$
at ${\cal O}(q^4)$ including the axial-vector meson $a_1(1260)$ explicitly. Full line: result with axial-vector
meson; dashed line: result without axial-vector meson.
   One can clearly see the dominant pion pole contribution at $Q^2\approx
   -M_\pi^2$.}
\end{figure}

   In Ref.\ \cite{Schindler:2006it} the form factors $G_A$ and
$G_P$ have been calculated in BChPT up to and including order ${\cal O}(q^4)$.
   In addition to the standard treatment including the
nucleon and pions, the axial-vector meson $a_1(1260)$ has also been considered as an explicit degree of freedom.
   The inclusion of the axial-vector meson effectively results in one additional
low-energy coupling constant which has been determined by a fit to the data for $G_A(Q^2)$.
   The inclusion of the axial-vector meson results in an improved
description of the experimental data for $G_A$ (see Fig.~\ref{H1_axff_GAwith}), while the contribution to $G_P$
is small.

\section{Complex-mass scheme and effective field theory}

   In Sec.\ \ref{subsec:FFVM} we saw how the inclusion of virtual vector mesons generates an
improved description of the electromagnetic form factors, for which ordinary chiral perturbation
theory does not produce sufficient curvature.
   So far the inclusion of virtual vector mesons has been restricted to low-energy processes
in which the vector mesons cannot be generated explicitly.
   However, one would also like to investigate the properties of hadronic resonances such as
their masses and widths as well as their electromagnetic properties.
   An extension of chiral effective field theory to the momentum region near the complex
pole corresponding to the vector mesons was proposed in Ref.~\cite{Djukanovic:2009zn},
in which the power-counting problem was addressed by applying the complex-mass scheme (CMS)
\cite{Stuart:1990,Denner:1999gp,Denner:2005fg,Denner:2006ic}
to the effective field theory.
   Since the $\rho$ mass is not treated as a small quantity, the presence of large external
four-momenta, e.g., in terms of the zeroth component, leads to a considerable complication
regarding the power counting of loop diagrams.
   To assign a chiral order to a given diagram it is first necessary to investigate all
possibilities how the external momenta could flow through the internal lines of that diagram.
   Next, when assigning powers to propagators and vertices, one needs to determine the
chiral order for a given flow of external momenta.
   Finally, the smallest order resulting from the various assignments is defined
as the chiral order of the given diagram.
   The application of the CMS to the renormalization of loop diagrams amounts to splitting
the bare parameters of the Lagrangian into renormalized parameters and counter terms and
choosing the renormalized masses as the {\it complex} poles of the dressed propagators
in the chiral limit, $M_R^2=(M_\chi-i\Gamma_\chi/2)^2$.
   The result for the chiral expansion of the pole mass and the width of
the $\rho$ meson to ${\cal O}(q^4)$ reads \cite{Djukanovic:2009zn}
\begin{eqnarray}
M_\rho^2  & = & M_\chi^2 +c_x M^2
-\frac{g_{\omega\rho\pi}^2 M^3 M_\chi}{24 \pi }\nonumber\\
&&+\frac{M^4}{32\pi^2 F^2} \left(3 -2\, \ln \frac{M^2}{M_\chi^2}\right) +\frac{g_{\omega\rho\pi}^2 M^4}{32 \pi
^2} \left(1-\ln \frac{M^2}{M_\chi^2}\right)\,,
\label{phmass}\\
\Gamma & = & \Gamma_\chi +\frac{\Gamma_\chi ^3}{8 M_\chi^2}-\frac{c_x \Gamma_\chi  M^2}{2 M_\chi^2}
-\frac{g_{\omega\rho\pi}^2 M^3 \Gamma_\chi}{48 \pi \,M_\chi} +\frac{ M^4}{16\,\pi \,F^2 M_\chi}\,.
\label{phwidth}
\end{eqnarray}
  Here, $M^2$ is the lowest-order expression for the squared pion mass, $F$ the pion-decay constant in the
chiral limit, $c_x$ a low-energy coupling constant of the $\pi\rho$ Lagrangian, and $g_{\omega\rho\pi}$
a coupling constant.
   The nonanalytic terms of Eq.~(\ref{phmass}) agree with the results of Ref.~\cite{Leinweber:2001ac}.
   Both mass $M_\chi$ and width $\Gamma_\chi$ in the chiral limit are input parameters in this approach. The numerical importance of the different contributions has been estimated using
$$
F=0.092 \,{\rm GeV},\quad M=0.139 \,{\rm GeV}\,,\quad g_{\omega\rho\pi} = 16 \,{\rm GeV^{-1}},\quad M_\chi\approx
M_\rho=0.78\,{\rm GeV},
$$
resulting in the expansion (units of GeV$^2$ and GeV, respectively)
\begin{eqnarray}
M_\rho^2 & = & M_\chi^2+0.019 \,c_x - 0.0071 + 0.0014 +0.0013\,,\nonumber\\
\Gamma & \approx & \Gamma_\chi + 0.21\,\Gamma_\chi^3-0.016\,c_x \Gamma_\chi -0.0058\,\Gamma_\chi + 0.0011\,.
\label{numestimate}
\end{eqnarray}
For pion masses larger than $M_\rho/2$ the $\rho$ meson becomes a stable particle. For such values of the pion
mass the series of Eq.~(\ref{phwidth}) should diverge. Along similar lines, Ref.~\cite{Djukanovic:2009gt}
contains a calculation of the mass and the width of the Roper resonance using the CMS.

\section{Conclusion}
   In the baryonic sector new renormalization conditions have reconciled
the manifestly Lorentz-invariant approach with the standard power counting.
   We have discussed some results of a two-loop calculation of
the nucleon mass.
   The inclusion of vector and axial-vector mesons as explicit
degrees of freedom leads to an improved phenomenological description of the electromagnetic and axial form
factors, respectively.
   Work on the application to electromagnetic processes such as Compton
scattering and pion production is in progress.
   When describing resonances in perturbation theory, one needs to take their finite
widths into account.
   The CMS has opened a new window for describing unstable particles in EFT with a consistent
power counting.

\acknowledgments
This work was made possible by the financial support from the
Deutsche Forschungsgemeinschaft (SFB 443 and SCHE 459/2-1).

\end{document}